\documentclass[aps,pre,preprint,showpacs,amsmath,amsfonts,preprintnumbers]{revtex4}
\usepackage{epsfig}
\usepackage{yfonts}
\usepackage{amssymb}
\usepackage{graphicx}
\swabfamily
\newcommand{\Op}[1]{{{\mathrm{\hat{#1}}}}}

\begin{document}
\title{Efficient simulation of  quantum evolution using dynamical coarse-graining.}
\author{M. Khasin and R. Kosloff}
\affiliation{Fritz Haber Research Center for Molecular Dynamics, 
Hebrew University of Jerusalem, Jerusalem 91904, Israel}
\date{\today }

\begin{abstract}
A novel scheme to simulate the evolution of a restricted set of observables of a quantum system is proposed.
The set comprises the spectrum-generating algebra of the Hamiltonian. Focusing on the   simulation of the restricted set of observables  allows to drastically reduce the complexity of the simulation.  This reduction is the result of replacing the original unitary dynamics
by a special open-system evolution. This open-system evolution can be interpreted as a process of weak measurement of the distinguished observables performed on the evolving system of interest. Under the condition that the observables are  "classical"  and the Hamiltonian is moderately nonlinear,  the open system dynamics displays a large time-scales separation between the relaxation of the observables and the decoherence of a generic  evolving state. 
The time scale separation allows the \textit{unitary} dynamics of the observables to be efficiently simulated by the \textit{open-system} dynamics on the intermediate time-scale. The simulation employs unraveling of the corresponding master equations into pure state evolutions, governed by the stochastic nonlinear Schr\"odinger equation (sNLSE). The stochastic pure state evolution can be efficiently simulated using a representation of the state in the time-dependent basis of the generalized coherent states, associated with the spectrum-generating algebra. 

\end{abstract}
\maketitle

\section{introduction}
The number of independent observables of a quantum system with the Hilbert space dimension $N$ is $N^2-1$.  
In many-body systems, when $N$ increases exponentially with the number of degrees of freedom, that large number of observables can be neither measured nor calculated.  Only a limited number of dynamical variables is accessible to an experimentalist, while all the uncontrollable parameters are averaged out. This means that generically, an observed quantum system is characterized by a small number of the expectation values of  accessible observables.  To theoretically characterize the dynamics of a quantum system it is desirable: (i) to find equations of motion for this reduced set of expectation values; (ii) to be able to solve the associated equations of motion efficiently. A computational cost of a direct quantum simulation scales  as $O(N^{\delta})$,  $\delta>1$ \cite{k56}.  A simulation is defined as efficient if the corresponding equations of motion can be solved with a computational cost which is reduced substantially  from that number.

In the present study we explore the possibility of  an efficient simulation of a restricted set of observables, using a novel paradigm for the simulation.  Assuming that the set of experimentally accessible observables is small, it is plausible that there exist a number of microscopic theories, leading to the same observed dynamics. If a microscopic theory can be found, which leads to equations of motion that can be solved efficiently, the dynamics of the restricted set of observables can be efficiently simulated.
More specifically, we propose to simulate the unitary dynamics of a quantum system by embedding it
in a particular \textit{open system} dynamics. In this dynamics the coupling to the bath is constructed to have a negligible impact on the evolution of the selected set of observables on the characteristic time scale of their \textit{unitary} evolution. The key point is that the resulting {open} system dynamics can  be efficiently simulated. 
The reduction of the computational complexity of the evolution, imposed by the bath, is attributed to dynamical coarse-graining, collapsing the system to a preselected representation which is used as the basis for the dynamical description. Since the bath has no observable effect by construction it should be considered solely as a computational tool. For that reason a term \textit{fictitious bath} is used in the paper to refer to it.

The quantum systems considered in the present work have finite Hilbert space dimension.
The dynamics is generated by the Lie-algebraic Hamiltonians:
\begin{eqnarray}
\Op H=\sum_i a_i \Op X_i+\sum_{ij} b_{ij}\Op X_i \Op X_j, \label{laham}
\end{eqnarray}
where the set $\{ \Op X_i \}$ of observables is closed under the commutation relations:
\begin{eqnarray}
\left[\Op X_i,\Op X_j\right]=i \sum_{k=1}^K  f_{ijk}\Op X_k ,\label{commutation}
\end{eqnarray}
i.e., it forms the spectrum-generating \cite{bohm} Lie algebra \cite{gilmorebook}  of the system. This algebra is labeled by the letter $\mathfrak g$ in what follows. Lie-algebraic Hamiltonians (\ref{laham}) are abundant in molecular \cite{Iachello, iachello06}, nuclear \cite{bohm, iachello06} and condensed matter physics \cite{bohm} . The basis of the algebra  $\{ \Op X_i \}$ is chosen as a distinguished set of observables, which are to be simulated efficiently.
Lie algebras considered in the present work are compact semisimple algebras \cite{gilmorebook} and the basis $\{ \Op X_i \}$ is assumed to be  orthonormal   with respect to the Killing form \cite{gilmorebook}. 

The corresponding open system dynamics, which is alleged to simulate the unitary dynamics of the elements of $\mathfrak g$, is governed by the following Liouville-von Neumann  equation of motion
\begin{eqnarray}
\frac{\partial}{\partial t}\Op \rho={\cal L} \Op \rho=-i \left[\Op H, \Op\rho \right]-\gamma \sum_{j=1}^K  \left[\Op X_j,\left[\Op X_j,\Op\rho \right]\right], \label{liouville}
\end{eqnarray} 
which has the  Lindblad form  \cite{lindblad76,breuer}, i.e., it describes a Markovian completely positive \cite{breuer} nonunitary evolution of the quantum system. The physical interpretation of the evolution, governed by Eq.(\ref{liouville}) is the process of weak measurements \cite{diosi06} of the algebra of observables $\mathfrak g$, performed on the quantum system, evolving under the Hamiltonian (\ref{laham}). 

The foundation of the method is  the observation that coupling to the bath induces a decoherence of the evolving density operator in a particular basis known as generalized coherent states (GCS), associated with the algebra (Section II). It is shown that if the Hamiltonian is linear in $ \Op X_i $ and a certain "classicality condition" is satisfied by the Hilbert space representation of the algebra, the decoherence time-scale is much shorter than the timescale on which the effect of the bath on the elements of $\mathfrak g$ is measurable, i.e., the relaxation time-scale. It is conjectured that this strong time separation will also hold for Hamiltonians bilinear in the elements of $\mathfrak g$ (Section III). 

We propose to take advantage of this property of the {open system dynamics} for efficient simulation of the \textit{unitary} evolution of $\{ \Op X_i \}$, using  stochastic unravelings of the evolution \cite{gisin84,diosi88,gisin92} and representing the evolving stochastic pure state in the time-depending basis of the GCS \cite{Perelomov,Gilmore} (Section IV). The effect of the decoherence translates into localization of evolving stochastic pure state in the GCS basis, which enables efficient representation and simulation of the stochastic evolution. Averaging over the unravelings recovers the {unitary} dynamics of the algebra generators. It is shown that the averaging can be performed efficiently provided the corresponding dynamics can be efficiently measured in the lab. The effect of coupling to the fictitious bath is illustrated by  the dynamics of a Bose-Einstein condensate (BEC) in a double-well trap \cite{schumm,gati} modeled by the two-mode Bose-Hubbard Hamiltonian (Section V). It is  demonstrated that the bath induces drastic localization on the level of a stochastic pure-state evolution, while having no observed effect on the dynamics of the elements of the  spectrum-generating algebra of the system.

\section{Evolution of states}

A central theme in this section is the intimate relation between the evolution of the subalgebra of observables and
the dynamics of the generalized coherent states (GCS) associated with this subalgebra.
The GCS  minimize the total uncertainty with respect to the  basis elements of the subalgebra and in addition
are maximally robust to interaction with the bath, modeled by Eq.(\ref{liouville}). 

\subsection{Generalized coherent states and the total uncertainty.}

Let us assume that the subalgebra $\mathfrak g$ is represented irreducibly on the system's Hilbert space $\cal H$. Then an arbitrary state $\psi\in \cal H$ can be represented as a superposition of the \textit{generalized coherent states} (GCS) \cite{Perelomov,Gilmore} $\left|\Omega,\psi_0\right\rangle$ with respect to the corresponding dynamical group $G$ and an arbitrary state $\psi_0$:
\begin{eqnarray}
\left|\psi\right\rangle=\int d \mu(\Omega)\left|\Omega,\psi_0\right\rangle\left\langle \Omega, \psi_0|\psi\right\rangle,  \label{expansion1}
\end{eqnarray}
where $\mu(\Omega)$ is the group invariant measure on the coset space $G/H$ \cite{gilmorebook} , $\Omega\in G/H$,  $H\subset G$ is the maximal stability subgroup of the reference state $\psi_0$:
\begin{eqnarray}
h\left|\psi_0\right\rangle=e^{i\phi(h)}\left|\psi_0\right\rangle, \ \ h\in H
\end{eqnarray}
and the GCS $\left|\Omega,\psi_0\right\rangle$ are defined as follows:
\begin{eqnarray}
\Op U(g)\left|\psi_0\right\rangle=\Op U(\Omega h)\left|\psi_0\right\rangle=e^{i\phi(h)}\Op U(\Omega ) \left|\psi_0\right\rangle\equiv e^{i\phi(h)} \left|\Omega, \psi_0\right\rangle, \ \ g\in G, \ h\in H, \ \Omega\in G/H, \label{gcs}
\end{eqnarray}
where $\Op U(g)$ is a unitary transformation generated by a group element $g\in G$.

The  group-invariant \textit{total uncertainty} of a state with respect to a compact semisimple algebra $\mathfrak g$ is defined as \cite{delbourgo, Perelomov}:
\begin{eqnarray}
\Delta(\psi)\equiv\sum_{j=1}^K \left\langle \Delta \Op X_j^2\right\rangle_{\psi}=\sum_{j=1}^K  \left\langle  \Op X_j^2\right\rangle_{\psi}- \sum_{j=1}^K \left\langle  \Op X_j\right\rangle_{\psi}^2\label{giu}.
\end{eqnarray}
The first term in the rhs of Eq.(\ref{giu}) is the eigenvalue  of the   the Casimir operator  of $\mathfrak g$ in the Hilbert space representation:
\begin{eqnarray}
\Op C=\sum_{j=1}^K \Op X^2_j \label{Casimir} 
\end{eqnarray}
and the second term is termed the generalized purity \cite{Viola03} of the state with respect to $\mathfrak g$:
\begin{eqnarray}
P_{\mathfrak g}[\psi]\equiv \sum_{j=1}^K \left\langle  \Op X_j\right\rangle_{\psi}^2 \label{gpurity}.
\end{eqnarray}
Let us define $\Delta_{min}$ as a minimal total uncertainty of a quantum state  and  $c_{\cal H}$ as the eigenvalue  of the   the Casimir operator  of $\mathfrak g$ in the system Hilbert space. Then
\begin{eqnarray}
\Delta_{min} \le \Delta(\psi)\le c_{\cal H}, \label{giu1}
\end{eqnarray}

The total uncertainty (\ref{giu}) is invariant under an arbitrary unitary transformation generated by $\mathfrak g$. Therefore,  all the GCS with respect to the subalgebra $\mathfrak g$ and a reference state $\psi_0$ have a fixed value of the total invariance. It has been proved in Ref.\cite{delbourgo} that 
the minimal total uncertainty $\Delta_{min}$ is obtained if and only if $\psi_0$ is a highest (or lowest) weight state of the representation (the Hilbert space). The value of  $\Delta_{min}$ is given by \cite{delbourgo, klyachko} 
\begin{eqnarray}
\Delta_{min}\equiv (\Lambda,\mu) \le \Delta(\psi)\le (\Lambda,\Lambda+\mu)=c_{\cal H}, \label{giu2}
\end{eqnarray}
where $\Lambda \in {\mathbb R}^r$ is the the highest weight of the representation, $\mu \in {\mathbb R}^r$ is the  sum of  the positive roots of $\mathfrak g$,  $r$ is the rank of $\mathfrak g$ \cite{gilmorebook} and $(,)$ is the Euclidean scalar product in ${\mathbb R}^r$.
The corresponding CGS were termed the generalized unentangled states with respect to the subalgebra $\mathfrak g$ \cite{Viola03, klyachko}. 
The maximal value  of the uncertainty is obtained in states termed maximally or completely entangled  \cite{Viola03, klyachko} with respect to $\mathfrak g$. The maximum value equals  $c_{\cal H}$ in the states having $\left\langle \psi\left|\Op X_j\right|\psi \right\rangle^2=0$ for all $i$. Such states exist in a generic irreducible representation of an arbitrary compact simple algebra of observables \cite{klyachko}. Generic superpositions of the GCS have larger uncertainty and are termed generalized entangled states with respect to $\mathfrak g$ \cite{Viola03, klyachko}.
In what follows, it is assumed that the reference state $\psi_0$  for the GCS minimize the total invariance (\ref{giu}). 

\subsection{Decoherence timescales.} 

The rate  of  purity loss in an arbitrary pure state $\Op \rho=\left|\psi\right\rangle\left\langle \psi\right|$ can be calculated  using Eq.(\ref{liouville}) as follows \cite{viola07}:
\begin{eqnarray}
 \frac{d}{dt}\texttt{Tr}\left\{ \Op \rho^2  \right\}&=&\texttt{Tr}\left\{2 \dot {\Op{\rho} }  \Op \rho \right\}=2 \texttt{Tr}\left\{ i \left[\Op H, \Op\rho \right]\Op \rho -\gamma \sum_{j=1}^K  \left[\Op X_j,\left[\Op X_j,\Op\rho \right]\right]\Op \rho   \right\} \nonumber \\
&=& -2 \gamma \texttt{Tr}\left\{ \sum_{j=1}^K  \left[\Op X_j,\left[\Op X_j,\Op\rho \right]\right]\Op \rho   \right\}=-4 \gamma  \sum_{j=1}^K \left( \left\langle \psi\left|\Op X_j^2\right|\psi \right\rangle-\left\langle \psi\left|\Op X_j\right|\psi \right\rangle^2\right)\nonumber \\
&=&-4\gamma \sum_{j=1}^K \left\langle \Delta \Op X_j^2\right\rangle_{\psi}, \label{rate}
\end{eqnarray}
i.e., the rate is proportional to the group-invariant uncertainty (\ref{giu}).
From Eqs.(\ref{rate}) and (\ref{giu1})  it follows that the time scale of the purity loss in a \textit{generic state} is $(\gamma c_{\cal H})^{-1}$, where $c_{\cal H}$ is the eigenvalue of the Casimir, Eq. (\ref{Casimir}). On the contrary, the rate of purity loss of a GCS  is determined by
$\Delta_{min}$, Eq.(\ref{giu2}), which implies that GCS are rubust against the influence of the bath  \cite{viola07}.

Assume that  
\begin{eqnarray}
\Delta_{min}\ll c_{\cal H}. \label{classicality}
\end{eqnarray}
The strong inequality (\ref{classicality}) can be interpreted as follows. Under the action of the bath, modeled by Eq.(\ref{liouville}),  a generic {superposition} of the GCS, Eq.(\ref{expansion1}), decoheres on the fast time scale $(\gamma c_{\cal H})^{-1}$ into a {proper mixture} of the GCS, which then follows the slow evolution on a time scale fixed by $\Delta_{min}$. As a consequence, the effect of the bath is to "diagonalize" the evolving density operator into a time dependent statistical mixture of the GCS.

Accordingly, $(\gamma c_{\cal H})^{-1}$ determines the \textit{decoherence time scale} of the density operator in the basis of the GCS.

Condition (\ref{classicality}) does not depend on the strength of coupling to the bath and therefore is a  property of   the subalgebra of observables and its  Hilbert space representation. Condition (\ref{classicality}) will be termed the \textit{classicality condition} on the algebra of observables 
 (see Appendix B for some examples).

\section{Evolution of the observables}
The purpose of the present section is to show that the classicality condition (\ref{classicality}) implies a large time-scales separation between the decoherence of the state and the relaxation of the observables comprising the spectrum-generating algebra of the system, if the Hamiltonian (\ref{laham}) is linear in the generators of the  algebra. It is conjectured that the time-scales separation is preserved by the Hamiltonians at most bilinear in the generators.
As a consequence,  the unitary evolution of the observables on the intermediate time scale can be simulated by the open system dynamics, while the effect of the decoherence can be employed for efficient simulation of the open dynamics.

Consider an Hamiltonian linear in the elements of the algebra $\mathfrak g$, i.e.,    all $b_{ij}=0$ in Eq. (\ref{laham}).
The corresponding Heisenberg equations for the observables in $\mathfrak g$ becomes:
\begin{eqnarray}
\frac{\partial}{\partial t}\Op X_i&=&-i \left[\Op H, \Op X_i \right]-\gamma \sum_{j=1}^K  \left[\Op X_j,\left[\Op X_j,\Op X_i \right]\right]\nonumber \\
&=&-i \sum_{k=1}^K  \left(i a_{ik} \right)\Op X_k-\gamma \sum_{j,l=1}^K  \left(i f_{jik}\right)\left( i f_{jkl}\right)\Op X_l\nonumber \\
&=&-i \sum_{k=1}^K   \left(i a_{ik} \right)\Op X_k-\gamma \sum_{j,l=1}^K  \left(T^j\right)^2_{il}\Op X_l, \label{Heisenberg}
\end{eqnarray} 
where $T^i_{jk}=i f_{ijk}$ is a matrix element of the adjoint representation  \cite{gilmorebook} of $\Op X_i$. It is assumed without loss of generality that $\mathfrak g$ is a compact \textit{simple} subalgebra of observables ( in the general case of a \textit{semisimple} algebra, the system of Eqs.(\ref{Heisenberg}) decouples into systems of equations for the simple components of the algebra). The coefficients in the r.h.s. of (\ref{Heisenberg}) obey
\begin{eqnarray}
\sum_{j=1}^K  \left(T^j\right)^2=C_2
\end{eqnarray}
where $C_2$ is the quadratic Casimir of $\mathfrak g$ in the adjoint representation. Therefore 
\begin{eqnarray}
\left(\sum_{j=1}^K  \left(T^j\right)^2\right)_{il}=(C_2)_{il}=c_{\texttt{adj}} \delta_{il} \label{casimir}
\end{eqnarray}
leading to
\begin{eqnarray}
\frac{\partial}{\partial t}\Op X_i&=&-i \sum_{k=1}^K \left(i a_{ik} \right)\Op X_k-\gamma c_{\texttt{adj}} \Op X_i, \label{Heisenberg1}
\end{eqnarray} 
which in a matrix notation reads
\begin{eqnarray}
\frac{\partial}{\partial t}\Op X &=&-i\left(A-\gamma c_{\texttt{adj}} \right)\Op X , \label{Heisenberg2}
\end{eqnarray}
where $A=A^{\dagger}$ is defined by $A_{kl}=ia_{kl}$ and $\Op X\equiv \{\Op X_1,\Op X_2,...\Op X_k\}$.
We define  $\Op Y\equiv \{\Op Y_1,\Op Y_2,...\Op Y_k\}$ by
\begin{eqnarray}
\frac{\partial}{\partial t}\Op Y_i &=&-i A \Op Y_i=-i \omega_i \Op Y_i, \label{ham}
\end{eqnarray}
where $\omega_i$ are real since $A$ is Hermitian.
Then $\Op Y$ diagonalize also Eq.(\ref{Heisenberg2}):
\begin{eqnarray}
\frac{\partial}{\partial t}\Op Y_i &=&\left(-i A -\gamma c_{{\texttt{adj}}}\right)\Op Y_i=\left(-i\omega_i-\gamma c_{\texttt{adj}}\right) \Op Y_i, \label{Heisenberg3}
\end{eqnarray}
leading to the solution of Eq.(\ref{Heisenberg2}):
\begin{eqnarray}
\Op Y_i(t) &=& \Op Y_i(0)e^{-\left(i\omega_i+\gamma c_{\texttt{adj}} \right)t}  \label{solution}
\end{eqnarray}
and 
\begin{eqnarray}
\Op X_i(t)=\sum_j c_{ij} \Op Y_i(t). 
\end{eqnarray}

The solution (\ref{solution}) is obtained for an arbitrary compact simple subalgebra of the system observables  $\mathfrak{g}\cong \mathfrak{su}(K)\subseteq \mathfrak{su}(N)$ for a quantum system in a $N$-dimensional Hilbert space.
It can be generalized to a  semisimple subalgebra of observables, i.e., a direct sum of simple subalgebras, $\mathfrak{g}=\oplus_{i=1}^n \mathfrak{su}(K_i)\subseteq \mathfrak{su}(N)$, corresponding to a tensor-product partition of the system Hilbert space ${\cal H}=\otimes_{i=1}^n {\cal H}_i$. In this case, Eq.(\ref{solution}) corresponds to local observables of any given subsystem.

From Eq.(\ref{solution}) we see that the expectation values of observables in  $\mathfrak g$ oscillate on the timescales $\omega_i$ and decay on the time scale $\gamma c_{\texttt{adj}}$. Consider an observable $\Op Y_i$ such that $\omega_i \gg \gamma c_{\texttt{adj}}$. When the measurement of $\Op Y_i$ in a time interval 
\begin{eqnarray}
(\omega_i )^{-1}\ll \tau \ll (\gamma c_{\texttt{adj}})^{-1}, \label{interval}
\end{eqnarray}
is performed,  the nonunitary character of the evolution cannot be discovered. Therefore,   given  the time interval $\tau$ any $\gamma$ with the property $\tau \ll (\gamma c_{\texttt{adj}})^{-1}$ will lead to apparently unitary  dynamics of $\Op Y_i$ on the time interval $\tau$ .

Next we note that since  $(\Lambda,\mu)\neq 0$ in Eq.(\ref{giu2}) (a positive root has strictly positive scalar product with the maximal weight vector) strong inequality (\ref{classicality}) implies $|\Lambda| \gg |\mu|$, which leads to the following strong inequality
\begin{eqnarray}
\sqrt{c_{\cal H}}\gg \sqrt{c_{\texttt{adj}}}. \label{casinequality}
\end{eqnarray}

Therefore, a time interval $\tau$  exists such that 
\begin{eqnarray}
(\gamma c_{\cal H} )^{-1}\ll \omega_i ^{-1}\ll \tau \ll (\gamma c_{\texttt{adj}})^{-1}. \label{interval2}
\end{eqnarray}
for some $i$ corresponding to an observable $\Op Y_i$ in Eq.(\ref{solution}).
The lhs of the inequality (\ref{interval2}) is the decoherence rate of a generic superposition of the GCS with respect to the algebra $\mathfrak g$ and the lhs is the decay rate of the observable $\Op Y_i$. This system of strong inequalities implies two important properties of the open system dynamics,  Eqs.(\ref{Heisenberg}): (i) a generic superposition of the GCS collapses into a mixture of the GCS on a time scale much shorter than a physically interesting time scale of the unitary evolution of the observable; (ii) the time scale of the unitary evolution of the observable is much shorter than its relaxation time scale.

It should be to emphasized that the classicality condition (\ref{classicality}) is not sufficient to imply properties (i) and (ii) in a generic case of nonlinear Hamiltonians  (\ref{laham}). Generic nonlinearity in the Hamiltonian is expected to introduce faster time-scales in the relaxation of the observables. The crucial question is whether these timescales are as short as the decoherence timescale $(\gamma c_{\cal H} )^{-1}$. Since the decoherence time-scale  depends not only on the algebra but also on its Hilbert space representation, it seems that in order to introduce as fast time-scales into the relaxation, the nonlinearity should  effectively couple most of the Hilbert-Schmidt basis of the system operators on the  physically interesting time-scale of the unitary evolution of the algebra elements. It is conjectured that a Hamiltonians at most bilinear in the elements of the spectrum-generating algebra does not possess such strong coupling property. A closer investigation of this important question is left to future research.

\section{Efficient simulation of the evolution of the spectrum-generating algebra of observables.}
\textit{Efficient simulation} is defined as a simulation based on a numerical solution of the first order  differential equations for  a  number of dynamical variables  which is
much smaller than the Hilbert space dimension of the system. 
 The number of dynamical variables $m$ cannot be smaller than the number of observables to be simulated, which  equals  the dimension $K$ of the spectrum-generating algebra $\mathfrak g$. If there is a large gap between the dimension
of the algebra  and the Hilbert space dimension $K=\texttt{dim}\{\mathfrak g\} \ll \texttt{dim}\{{\cal H}\}=N$ the simulation based on the the number of variables
$K \lesssim  m \ll N$ is  considered efficient.

The principle of efficient simulation of the observables, forming the spectrum-generating algebra $\mathfrak g$ of the Hamiltonian (\ref{laham}) is based upon
\begin{itemize}
\item{Simulating the unitary evolution of the observables by the fictitious open system dynamics, governed by the Liouville-von Neumann Eq.(\ref{liouville});}
\item{Unraveling the Liouville-von Neumann Eq.(\ref{liouville}) into  {\textit{pure state}} evolutions, governed by the stochastic nonlinear Schr\"odinger equation (sNLSE) (see below);}
\item{Efficient simulation of the stochastic nonlinear pure state dynamics, using expansion of the state in a time-dependent basis of the generalized coherent states (GCS), associated with the spectrum-generating algebra $\mathfrak g$. }
\end{itemize}
In the previous section we have discussed the first of the listed items. The other two items focus on the principles of efficient simulation of the open-system evolution.

Solving directly the Liouville-von Neumann master Eq.(\ref{liouville}) is more difficult than the original problem. A  reduction in complexity is based on the equivalence between the Liouville-von Neumann
equation and the stochastic nonlinear Schr\"odinger equation (sNLSE)\cite{gisin84,diosi88,gisin92}: 
\begin{eqnarray}
d\left|{\psi}\right\rangle &=&\left\{-i \Op H dt -\gamma \sum_{i=1}^K   \left( \Op X_i-\left\langle \Op X_i\right\rangle_{\psi}  \right)^2 dt+ \sum_{i=1}^K\left(  \Op X_i-\left\langle \Op X_i\right\rangle_{\psi}\right) d\xi_i\right\}\left|{\psi}\right\rangle, \label{snlse}
\end{eqnarray}
where the Wiener fluctuation terms $d\xi_i$ satisfy 
\begin{eqnarray}
<d\xi_i>=0, \ \ \ d\xi_id\xi_j=2\gamma dt.
\end{eqnarray}
To demonstrate the equivalence, Eq.(\ref{snlse}) can be cast into the evolution 
of the projector $\Op P_{\psi}=\left|\psi\right\rangle\left\langle \psi\right|$ 
\begin{eqnarray}
d \Op P_{\psi}=\left(-i \left[\Op H, \Op P_{\psi} \right]-\gamma \sum_{j=1}^K  \left[\Op X_j,\left[\Op X_j,\Op P_{\psi} \right]\right]\right)dt+\sum_i  \left\{\left( \Op X_i-\left\langle \Op X_i\right\rangle_{\psi}   \right)d \xi_i,\Op P_{\psi}\right\}. \label{stochliouville}
\end{eqnarray}  
Averaging Eq.(\ref{stochliouville}) over the noise recoveres the original Liouville-von Neumann equation (\ref{liouville}). Therefore, the problem of efficient simulation of the Liouville-von Neumann dynamics
is transformed to the problem of efficient simulation of the nonlinear stochastic dynamics, governed by sNLSE Eq. (\ref{snlse}). 

The simulation of   the pure state evolution according to the sNLSE(\ref{snlse}) is based on an expansion of the evolving state in the time-dependent basis of the GCS, Eq.(\ref{expansion1}). In the case of a finite Hilbert space an arbitrary state can be represented as a superposition of $M \le N$ GCS:
\begin{eqnarray}
\left|\psi\right\rangle=\sum_{i=1}^M c_i\left|\Omega_i,\Lambda \right\rangle, \label{expansion2}
\end{eqnarray}
where $\Omega_i$ is an element of the coset space $G/H$, $G$ is the dynamical group of the system generated by $\mathfrak g$,  $H$ is  the maximal stability subgroup, corresponding to the reference state $\left|\Lambda \right\rangle$ and $\Lambda$ is the highest weight of the Hilbert space representation of the algebra. The coset space $G/H$ has natural symplectic structure \cite{Gilmore} and  can be considered as a phase space of the quantum system, corresponding to $\mathfrak g$. Accordingly, $\Omega_i$ is a point in the  phase space. The total number of variables defining (up to an overall phase) the state $\psi$ (\ref{expansion2}) equals $M$ times the dimension of the phase space $G/H$ plus the number $M$ of amplitudes $c_i$. The dimension of $G/H$ depends on the properties of the Hilbert space representation of the algebra, but is always strictly less then the dimension of the algebra $K$ \cite{Gilmore}. Therefore, the number $m$ of real parameters, characterizing the state $\psi$ (\ref{expansion2}) satisfies the following inequality
\begin{eqnarray}
m<M (K+2). \label{params}
\end{eqnarray}
It follows that the necessary condition for efficient simulation of the dynamics is that $1 \lesssim  M \ll N$ in the physically relevant time interval.

It is assumed that initial state of the system is a GCS, corresponding to $M=1$ in the expansion (\ref{expansion2}). If one omits the  nonlinear and stochastic terms in the Eq.(\ref{snlse}) it becomes a regular Schr\"oedinger equation, governing the unitary evolution of the state.
If the Hamiltonian in Eq.(\ref{snlse}) is linear in the elements of $\mathfrak g$, the initial GCS evolves into a GCS by the definition, Eq.(\ref{gcs}). Restoring the nonlinear and stochastic terms to Eq.(\ref{snlse}) 
breaks the unitarity of the evolution but  a GCS still evolves into a GCS 
under the full equation, Ref.\cite{khasin082}. Therefore a GCS solves the sNLSE(\ref{snlse}), driven by a linear Hamiltonian. In Ref.\cite{khasin082} it is proved that a CGS is a globally stable solution in that case, i.e., an arbitrary initial state evolves asymptotically into a GCS.

Adding bilinear terms to the Hamiltonian (\ref{laham}) breaks the invariance of the subalgebra $\mathfrak g$ under the action of the Hamiltonian and, as a consequence, an initial GCS evolves into a \textit{superposition} of a number $M>1$ of the GCS (\ref{expansion2}) in the corresponding unitary evolution. If the number of terms $M$ becomes large, $M=O(N)$, the unitary evolution can no longer be simulated efficiently. The nonlinear and stochastic terms (representing the effect of the fictitious bath) in Eq.(\ref{snlse}) is expected to  \textit{decrease} the effective number $M$ of terms in the expansion (\ref{expansion2}) of the evolving state. This effect will be termed \textit{localization}. The natural measure of the localization is the total uncertainty of the evolving state with respect the spectrum-generating algebra  $\mathfrak g$ or, equivalently, the generalized purity of the state with respect to $\mathfrak g$ \cite{violabrown}.

The localizing effect of the bath is proved in the Ref.\cite{khasin082}. Heuristically, it can be understood as follows. If each sum in the sNLSE (\ref{snlse}) is replaced by  a single contribution of a given operator $\Op X$  the uncertainty of the evolving state with respect to $\Op X$ is strictly decreasing under the action of the bath, unless the state is an eigenstate of $\Op X$, in which case it vanishes \cite{gisin84, diosi88, gisin92}. Therefore, the effect of the bath is to bring an arbitrary state into an eigenstate of $\Op X$. In our case, the observables $\Op X_i$ are noncommuting and cannot be diagonalized simultaneously. Therefore, it is expected that the effect of the bath in this case will be to take an arbitrary state to the state which minimizes the total uncertainty with respect to the elements of the algebra, i.e., to a GCS. 

The characteristic time scale of the localization is the decoherence time scale $(\gamma c_{\cal H} )^{-1}$. If  the classicality condition (\ref{classicality}) holds and  the nonlinearity of the Hamiltonian is moderate (Cf. the end of Section III), the localization is effective on a  time interval much shorter than the relaxation of the observables in $\mathfrak g$. As a consequence,   the unitary dynamics of these observables can be obtained by (i) simulating the nonlinear stochastic evolution of the localized pure states, (ii) calculating the  expectation values of the observables in each stochastic unraveling and (iii) averaging over the stochastic realizations.

Calculating the expectation values and averaging (steps (ii) and (iii) above) are not part of the definition of efficient simulation, and therefore should be considered separately. 
Even if the step (i) can be performed efficiently according to the definition, it is left to show that the complexity of performing steps (ii) and (iii), measured, for example, by a number of elementary computer operations, scale substantially less than the size of  the Hilbert space dimension.

To calculate the expectation value of an observable in a state represented by the GCS expansion (\ref{expansion2}) one has to calculate $M(M+1)/2$ matrix elements of the operator between the GCS. Each matrix element for an operator $\Op X_i\in \mathfrak g$ can be calculated group-theoretically \cite{Gilmore,Somma}, i.e., independently on the Hilbert space  representation. Therefore, if $M \ll N$ the computation of the expectation values of the elements of $\mathfrak g$  can be performed efficiently.

Complexity of the step (iii) is measured by the number of stochastic realizations necessary to obtain the expectation values of the observables to a prescribed accuracy. In Appendix C it is shown that the averaging  can be performed efficiently, provided the expectation values of the elements of  $\mathfrak g$ can be measured efficiently. More precisely
\begin{eqnarray}
n_{ st}(\epsilon) \le n_{ ex}(\epsilon)\texttt{dim}\{ \mathfrak g \},
\end{eqnarray}
where $n_{ st}(\epsilon)$ is the number of stochastic realizations, necessary to obtain the expectation value of each observable $\Op X_i \in \mathfrak g$ to an absolute accuracy $\epsilon$, $n_{ ex}(\epsilon)$ is the number of experimental runs, necessary to obtain the expectation value of each $\Op X_i$ to the absolute accuracy $\epsilon$ and $\texttt{dim}\{ \mathfrak g \}$ is the dimension of the subalgebra of observables which is assumed to be a small number. It follows that the averaging can be done efficiently, provided the measurement can be performed efficiently, which is assumed.
 In addition, it is important to emphasize that it is not necessary to converge the averaging process in order to obtain a meaningful information: even a single "trajectory" bears important information. 

We shall finally focus on the  step (i) of simulating the nonlinear stochastic evolution of the localized pure states. The localization means that the number of GCS terms $M$ in the expansion (\ref{expansion2}) is much smaller than the Hilbert space dimension $N$ and, by  virtue of the inequality (\ref{params}), the number $m$ of parameters that characterize the evolving state is much smaller than $N$.

The details of the derivation of equations of motion for the parameters will be given elsewhere \cite{preparation}. Here we point out the main ingredients of the derivation.
We put the sNLSE (\ref{snlse}) in the equivalent exponential form
\begin{eqnarray}
|\psi>+|d\psi>&=&\exp\left\{-i \Op H dt -\gamma \sum_{i=1}^K   \left( \Op X_i-\left\langle \Op X_i\right\rangle_{\psi}  \right)^2 dt+ \sum_i\left(  \Op X_i-\left\langle \Op X_i\right\rangle_{\psi}\right) d\xi_i\right\}|\psi> \nonumber \\ \label{infi1}\\
&=&\exp\left\{-\gamma \sum_{i=1}^K   \left( \Op X_i-\left\langle \Op X_i\right\rangle_{\psi}  \right)^2  dt+ \sum_i\left(  \Op X_i-\left\langle \Op X_i\right\rangle_{\psi}\right) d\xi_i\right\}e^{-i \Op H dt}|\psi>, \nonumber
\end{eqnarray} 
using the fact that the infinitesimal transformations commute to the leading order. 

The transformation
\begin{eqnarray}
|\psi'>=e^{-i \Op H dt}|\psi> \label{unitary}
\end{eqnarray}
is a unitary evolution, corresponding to the Schr\"oedinger equation. The first order differential equation of motions of parameters of the representation (\ref{expansion2}) under this unitary evolution can be derived variationally \cite{kramer}, using (\ref{expansion2}) as a variational ansatz. Therefore, the unitary evolution can be simulated efficiently, provided the number of terms in the expansion (\ref{expansion2}) is small.

Consider the second, nonunitary transformation 
\begin{eqnarray}
|\psi'>&=&\exp\left\{-\gamma \sum_{i=1}^K   \left( \Op X_i-\left\langle \Op X_i\right\rangle_{\psi}  \right)^2  dt+ \sum_{i=1}^K\left(  \Op X_i-\left\langle \Op X_i\right\rangle_{\psi}\right) d\xi_i\right\}|\psi'>\nonumber \\
&\stackrel{*}=&e^{\phi(t)}\exp\left\{ \sum_{i=1}^K \Op X_i\left(2 \gamma\left\langle \Op X_i\right\rangle_{\psi}  dt+  d\xi_i\right)\right\}|\psi'>\nonumber \\
&=&e^{\phi(t)} \sum_{i=1}^M c'_i\exp\left\{ \sum_{i=1}^K \Op X_i\left(2 \gamma\left\langle \Op X_i\right\rangle_{\psi}  dt+  d\xi_i\right)\right\} \left|\Omega'_i,\Lambda \right\rangle \nonumber \\
&\stackrel{**}=&e^{\phi(t)}\sum_{i=1}^M c'_i e^{\phi_i}\left|\Omega''_i,\Lambda \right\rangle=\sum_{i=1}^M c''_i \left|\Omega''_i,\Lambda \right\rangle, \label{nonunitary}
\end{eqnarray}
where  the starred equality follows from the fact that the Casimir operator $\sum_{i=1}^K \Op X_i^2 $ act as identity on an arbitrary state $\psi'$, and 
the double-starred equality follows from the fact that a not-necessarily-unitary transformation generated by an element of the algebra maps a GCS to a GCS modulo a complex phase \cite{Gilmore}. This transformation can be performed group-theoretically \cite{Gilmore}, i.e., efficiently.

The unitary evolution, Eq.(\ref{unitary}), generated by the nonlinear Hamiltonian (\ref{laham}), will lead to delocalization of the evolving state. The nonunitary evolution, Eq.(\ref{nonunitary}), $-$ to localization. At sufficiently strong localization  the number of terms $M$ necessary to converge the solution of the sNLSE (\ref{snlse}) on a fixed time interval will be much smaller, than in the corresponding unitary evolution, and efficient simulation of sNLSE (\ref{snlse}) will become feasible.

  The next section takes up an  example of a two-mode  Bose-Hubbard model of a Bose-Einstein condensate in a double-well trap to  illustrate the localizing properties of the fictitious  bath.

\section{Example: Two-mode Bose-Hubbard model.}

A common model for an ultracold gas of bosonic atoms in a one-dimensional periodic optical lattice
is described by the Bose-Hubbard Hamiltonian: 
\begin{eqnarray}
\Op H=-\Delta \sum_i (\Op a_{i+1}^{\dagger} \Op a_i+\Op a_{i}^{\dagger} \Op a_{i+1})+\frac{U}{2}\sum_i (\Op a_{i}^{\dagger} \Op a_i)^2 \label{bhgenral},
\end{eqnarray}
where $\Delta$ is the nearest neighbors hopping rate and $U$ is the strength of the on-site interactions
between particles. 
In the simplest case of a two-sites lattice model, which has been realized experimentally by confining a condensate in a double-well trap \cite{schumm,gati}, the Hamiltonian (\ref{bhgenral}) reduces to
\begin{eqnarray}
\Op H=-\Delta (\Op a_{1}^{\dagger} \Op a_2+\Op a_{2}^{\dagger} \Op a_{1})+\frac{U}{2}\left((\Op a_{1}^{\dagger} \Op a_1)^2+(\Op a_{2}^{\dagger} \Op a_2)^2\right) \label{bhdouble},
\label{eq:twosite}
\end{eqnarray}
where $\Delta$ is the tunneling rate.
Eq. (\ref{eq:twosite}) can be transformed \cite{vardi00} to the ${\mathfrak{su}}(2)$ set of operators
\begin{eqnarray}
\Op J_x&=&\frac{1}{2}( \Op a_{1}^{\dagger} \Op a_2+\Op a_{2}^{\dagger} \Op a_{1} ) \nonumber \\
\Op J_y&=&\frac{1}{2 i}( \Op a_{1}^{\dagger} \Op a_2-\Op a_{2}^{\dagger} \Op a_{1} ) \label{suops} \\ 
\Op J_z&=&\frac{1}{2}( \Op a_{1}^{\dagger} \Op a_1-\Op a_{2}^{\dagger} \Op a_{2} ) \nonumber
\end{eqnarray}
leading to the following Lie-algebraic form
\begin{eqnarray}
\Op H=-\omega \Op J_x+ U \Op J_z^2 \label{bhlie},
\end{eqnarray}
where $\omega=2\Delta$.
The Hilbert space of the system of $N$ bosons in this model corresponds to the $j=N/2$ irreducible representation of the ${\mathfrak{su}}(2)$ algebra.
We seek to simulate the evolution of the operators (\ref{suops}), driven by the Hamiltonian (\ref{bhlie}), where the initial state of the system is a GCS with respect to the ${\mathfrak{su}}(2)$, the spin-coherent state \cite{arecchi,Gilmore,viera} . 
More specifically, the initial state is chosen as
\begin{eqnarray}
\left|\psi(0)\right\rangle=\left|-j\right\rangle \label{initial},
\end{eqnarray}
which corresponds to the state of the condensate, localized in a single well.

The dynamics driven by the weak measurement of the operators (\ref{suops}) on the evolving condensate is described by  the Liouville-von Neumann equation of the form (\ref{liouville}):
\begin{eqnarray}
\frac{\partial}{\partial t}\Op \rho=-i [\Op H,\Op \rho ]-\gamma\sum_{i=0}^2 [\Op J_i,[\Op J_i,\Op \rho]]. \label{bhbath}
\end{eqnarray}

The classicality condition (\ref{classicality}) for the $2j+1=N+1$-dimensional representation of the ${\mathfrak{su}}(2)$, corresponding to $N$ atoms in the trap, translates into the $N \gg1$ condition (Appendix B). Therefore, for sufficiently large numbers of atoms in the trap the classicality is satisfied and a sufficiently weak measurement of the operators $\Op J_x$, $\Op J_y$ and $\Op J_z$ is expected to induce strong decoherence in the spin-coherent state basis, but leaving the dynamics of the operators practically unperturbed. As a consequence, the generalized purity of a stochastic unraveling of the Eq.(\ref{bhbath}) $P_{{\mathfrak{su}}(2)}[\psi]=\sum_i \left\langle \Op J_i/j\right\rangle^2$ is expected to remain close to unity, which enables efficient simulation of the corresponding dynamics.

\begin{figure}[t]
\epsfig{file=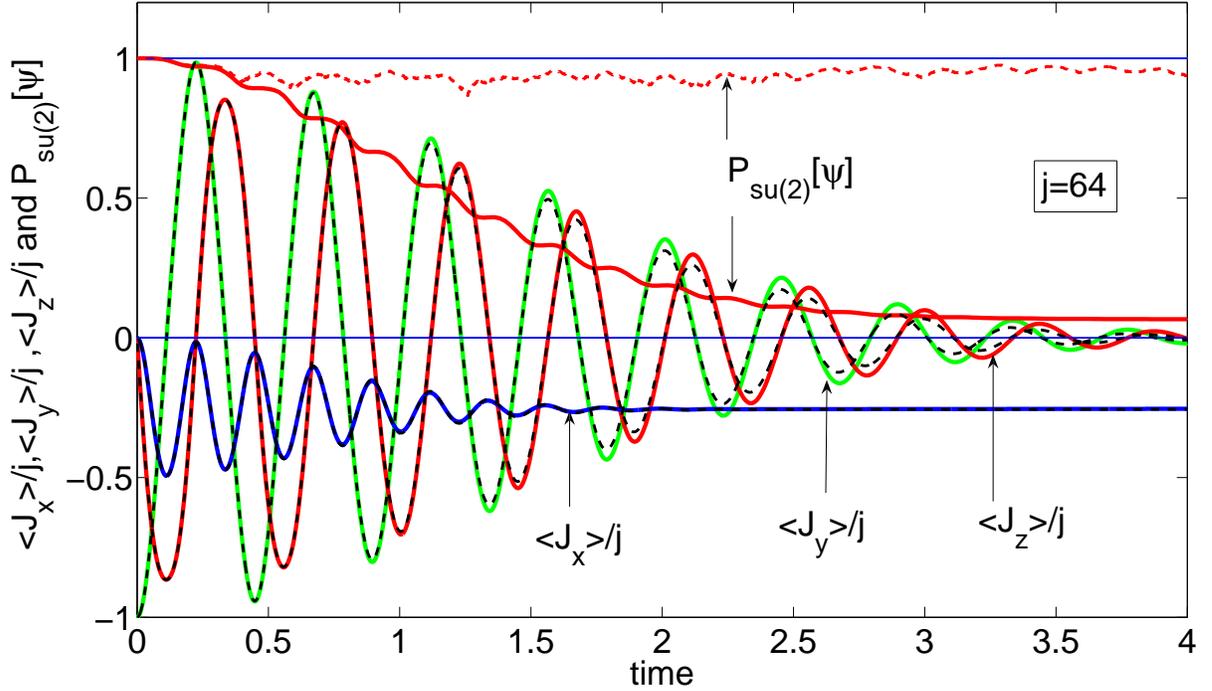, width=18.0cm, clip=} 
\caption{ The purity and expectation values of observables as a function of time.
An initial  GCS, Eq.(\ref{initial}),  undergoes (i) unitary, $\gamma=0$ (solid lines); (ii) nonunitary, $\gamma=0.05/j$ (dashed lines),  evolution according to the Liouville Eq. (\ref{bhbath}). The values of parameters chosen for the numerical solution are $\omega=15$ and $U=\omega/2j$. The observed dynamics of the expectation values of $\Op J_x/j$, $\Op J_y/j$ and $\Op J_z/j$ is negligibly affected by the bath while the generalized purity $P_{{\mathfrak{su}}(2)}[\psi]$ of the stochastic unraveling of the nonunitary evolution is larger by the factor of $15$ than the minimal purity of the unitarily evolving state.}
\label{fig:bathdriven}
\end{figure}
\begin{figure}[t]
\epsfig{file=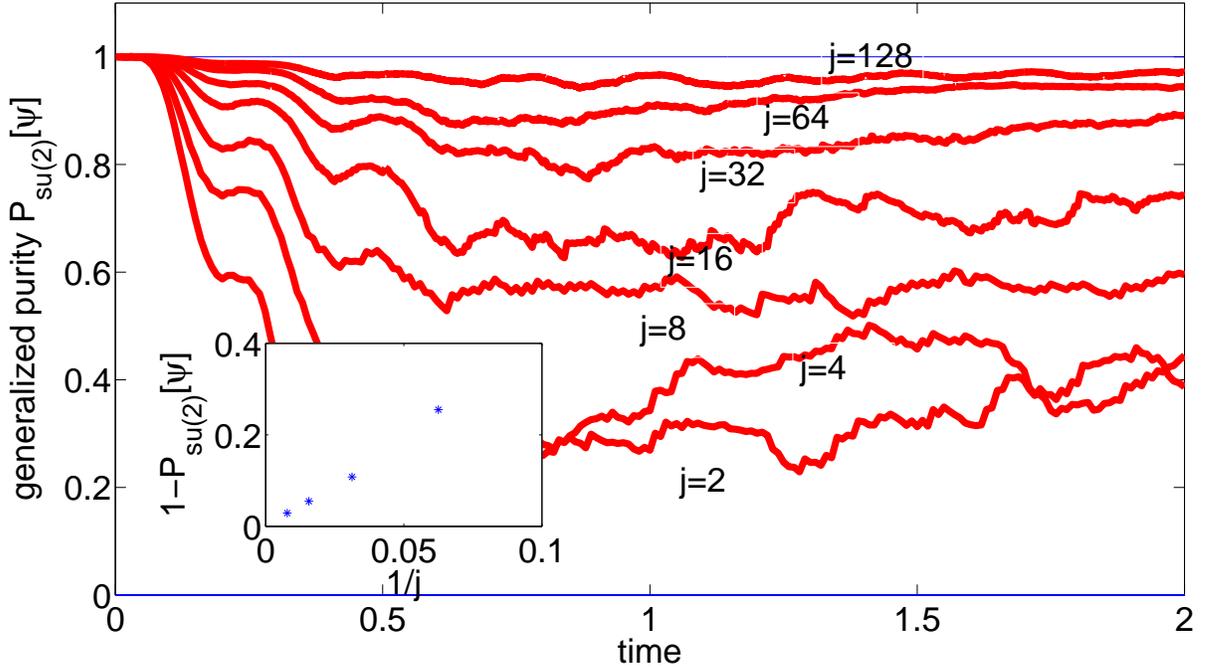, width=18.0cm, clip=} 
\caption{ Generalized purity  averaged over a small number (2-10) of stochastic unravelings of the Liouville-von Neumann Eq. (\ref{bhbath}). Initial state and parameters of the equation are as in the Fig. (1). Purity is plotted for $j=2,4,8,16,32,64,128$. The insert shows the generalized purity as a function of $1/j$.
At larger $j$ the value of the averaged purity is apparently consistent with the estimate $1-\frac{1}{j}f(\omega^{a} U^b \gamma^c)$, with $f(\omega^{a} U^b \gamma^c)=3$, corresponding to $M=3$ number of the GCS terms in the expansion of the solution.}
\label{fig:bathdriven2}
\end{figure}

Fig. (\ref{fig:bathdriven}) displays the evolution of the expectation values of the  operators $\Op J_x/j$, $\Op J_y/j$ and $\Op J_z/j$ in the unitary evolution $\gamma=0$  and in the nonunitary case $\gamma=0.05/j$ for $N=2j=128$ particles in the condensate. The hopping rate $\omega=15$ and the strength of the on-site interaction is $U=\omega/2j$. It can be seen that the evolution is negligibly perturbed by the bath for the chosen strength of the coupling $\gamma$. We also plot the generalized purity of the unitarily evolving state and of a random stochastic unraveling of the nonunitary evolution. The generalized purity in the unitary case decreases to the value of about 0.06, which corresponds (Appendix A) to the number of configurations $M=0.75 (2j+1)\approx 100=O(N)$ in the GCS expansion of the solution.  On the other hand, the generalized purity in the stochastic unraveling is about $0.9-0.95$ which corresponds to a drastic reduction of the number of configurations to $M=0.04(2j+1)\approx 5 \ll N$.

An interesting feature of the stochastic evolution displayed on Fig.(\ref{fig:bathdriven}) (and observed in  other numerical simulations, see Fig.(2)) is that apparently, the generalized purity approaches a constant value on average. Since the generalized purity is a measure of localization of the state on the corresponding phase space (which is the Bloch sphere for the ${\mathfrak {su}}(2)$ algebra \cite{arecchi,Gilmore,viera}) such behavior is suggestive of a soliton-like solution of the sNLSE (\ref{snlse}).  Investigation of existence and properties of these soliton-like solutions seems to be an interesting topic for future research. For the time being let us assume  that the stationary (on average) value $P$ of the generalized purity as displayed on Fig.(\ref{fig:bathdriven}) is an analytical function of $1/j$ (see Fig.2 for some evidence). Then
\begin{eqnarray}
P=1-\frac{1}{j}f(\omega^{a} U^b \gamma^c),
\end{eqnarray}
to the lowest order in $1/j$, where $f$ is an unknown function of the dimensionless argument $\omega^{a} U^b \gamma^c$ and $a+b+c=0$. Using the estimate (Appendix A) for the number of configurations we obtain
\begin{eqnarray}
M=(2j+1)(1-\sqrt{P})=f(\omega^{a} U^b \gamma^c),
\end{eqnarray}
i.e., the number of configurations in the expansion of the stochastic unraveling does not depend on $j$. Numerical evidence  implies that generally $f(\omega^{a} U^b \gamma^c)\neq 1$. For example, the value of  $f(\omega^{a} U^b \gamma^c)$ deduced from the  Fig. 2 is {\em three}.  This implies, that asymptotically, as $j\rightarrow \infty$, the dynamics of the single-particle observables of the two-modes Bose-Hubbard model can be reproduced not by an averaging over stochastic \textit{GCS} evolutions (stochastic \textit{mean-field solutions}), but rather by an averaging over the stochastic evolutions of \textit{superpositions} of a constant small number $M>1$ of GCS. 

The current observations were also found in different parametric regimes of the Bose-Hubbard model.
Similar behavior has been observed in the study of the ${\mathfrak{su}}(2)$-Hamiltonians, such as the Lipkin-Meshkov-Glick model \cite{lmg} of a system of interacting fermions. 

\section{Discussion and Open Questions}

The novel strategy for efficient simulation of a unitary evolution of a restricted set of observables has been outlined (Cf. Fig.3). The restricted set comprises the spectrum-generating Lie-algebra of the system. The unitary evolution of the observables is simulated by a particular open-system dynamics, corresponding to the process of weak measurement of the observables, performed on the evolving quantum system. 
The scheme is based on a large time-scales separation between the decoherence of the evolving state in the basis of the generalized coherent states (GCS), associated with the algebra, and the relaxation of the elements of the algebra. The necessary condition for the successful implementation of the algorithm is 
 the "classicality" condition  on the spectrum-generating algebra and its Hilbert space representation. 
This time-scale separation is proved for linear Hamiltonians and it is conjectured that the time-scales separation is preserved if the Hamiltonian is at most bilinear in the elements of the algebra.  
Numerical evidence obtained in the ${\mathfrak{su}}(2)$  case supports the conjecture. 
The "classicality" condition  excludes efficient simulation of certain subalgebras of observables (Appendix B). For example, the unitary dynamics of local observables of a composite system of qubits cannot be simulated with higher efficiency by the open-system evolution.

\begin{figure}[t]
\epsfig{file=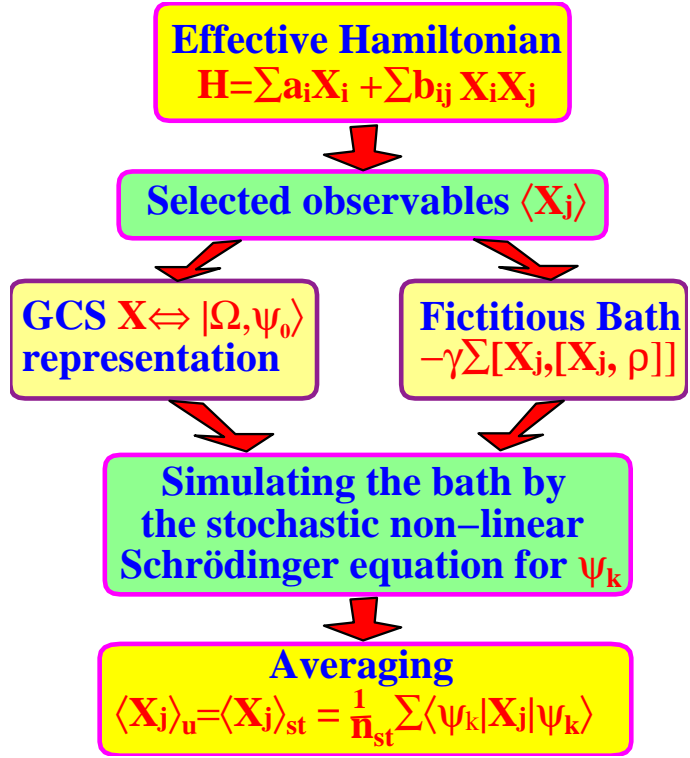, width=9.0cm, clip=} 
\caption{A schematic flow chart of the proposed approach to simulate dynamics of the operators $\Op X_i$ of the spectrum-generating algebra of the system. The unitary evolution of the observables is simulated by the open-system evolution, modeling  weak measurement of the evolving observables. The open-system dynamics is unraveled into stochastic pure-state evolutions, efficiently simulated using expansion of the pure state in the GCS. Efficient averaging over $n_{st}$ realizations obtains the expectation values, corresponding to the unitary evolution $\left\langle \Op X_j\right\rangle_{u}=\left\langle \Op X_j\right\rangle_{st}$.}
\label{fig:bathdriven3}
\end{figure}

The fast decoherence reduces the computational complexity of the evolution, while the slow relaxation leaves the dynamics of the restricted set of observables practically unaffected on physically interesting time-scales. The effect of a fictitious coupling to a bath can be viewed as a dynamically induced coarse-graining of the evolving state in the phase-space, associated with the spectrum-generating Lie-algebra. The fine structure of the evolving state, irrelevant for the expectation values of the "smooth" observables, is rubbed out by the decoherence, thereby reducing the computational complexity of the evolution. This coarse-graining can be seen as a generalization of the process of conversion of quantum correlations (entanglement) to classical correlations under the action of local dephasing environments \cite{khasin07}.
The reduction of the computational complexity is realized  by simulating the stochastic nonlinear Schr\"odinger equation (sNLSE), governing the stochastic unraveling of the nonunitary evolution. The  GCS  are globally stable solutions of the  sNLSE, corresponding to an Hamiltonian, linear in the algebra elements \cite{khasin082}.  Numerical evidence obtained in the ${\mathfrak{su}}(2)$  case suggests that  Hamiltonians bilinear in the generators asymptotically lead to soliton-like stable localized solutions of the corresponding sNLSE. Averaging over the stochastic realizations of the open-system evolution recovers the \textit{unitary dynamics} of the restricted set of observables. It is proved that the averaging can be performed efficiently, provided that the observables can be efficiently measured in an experiment.

The  fictitious bath is fine-tuned $-$ it corresponds to a process of weak measurement of the orthonormal basis set of the operators, performed with equal rates and strengths. This fine-tuned   bath is constructed  as a computational tool.  On the other hand, if the fine-tuning condition is dropped, the resulting open-system dynamics can represent a real physical situation, where the linear part of the Hamiltonian is perturbed by the time-dependent $\delta$-correlated noise  \cite{gorini76}. In that case the density operator of the system will follow an open evolution, corresponding to a process of weak measurement of the algebra elements, performed with generally different rates  \cite{gorini76}. It is expected, that if the noise is sufficiently weak, the constant part of the Hamiltonian will induce fast (on the relaxation time-scale) rotation in the Hilbert-Schmidt operator space, which effectively will average out the difference between the contributions of various measurements. Therefore, this real bath is expected to induce the same type of localization as the fine-tuned fictitious bath. Numerical evidence obtained in the ${\mathfrak{su}}(2)$  case supports this conjecture \cite{preparation}. Restricting the measurements to the algebra elements, the experimentalist will not observe the effect of the bath if the noise is sufficiently small, while measuring the higher order correlations will reveal the nonunitary character of the evolution. Generally, it is expected that the resulting open-system dynamics can be simulated  with higher efficiency than
the corresponding unitary dynamics, provided the "classicality" condition holds.

The main open questions  to be addressed are:
\begin{itemize}
\item{Investigation of the effect of nonlinear terms in the Hamiltonian (\ref{laham}) on the relaxation time-scales of the observables in the spectrum-generating algebra  in the corresponding fictitious open-system dynamics, Eq.(\ref{liouville}).  }
\item{Development of an efficient and convergent algorithm for simulating the evolution of a state in the GCS basis representation.}
\item{A rigorous proof of the conjecture, that for Lie-algebraic Hamiltonians, bilinear in the generators, Eq.(\ref{laham}),  the generalized purity of the stochastic unraveling of the corresponding open system evolution, Eq.(\ref{liouville}),  is stationary on average and the stationary purity approaches a limiting value independent on the dimension of the Hilbert space if the classicality condition becomes stronger.}
\end{itemize}

\begin{acknowledgments}
We are grateful to H. Barnum, L. Diosi, Y. Khodorkovsky, D. Steinitz and A. Vardi for discussions.
Work supported by DIP and the Israel Science Foundation (ISF).
The Fritz Haber Center is supported
by the Minerva Gesellschaft f\"{u}r die Forschung GmbH M\"{u}nchen, Germany.
\end{acknowledgments}
\appendix

\section{Relation of the generalized purity  to the number of  configurations in the GCS expansion of the state. ${\mathfrak{su}}(2)$ case.}

The phase space of a quantum system, associated with the ${\mathfrak{su}}(2)$ spectrum generating algebra is a two-dimensional sphere \cite{arecchi,Gilmore,viera}, usually called a Bloch sphere.
The localization of a state $\psi$ of the system  in the  phase space means localization of its $P$-distribution \cite{Gilmore,viera}  about a point in the phase space. Without loss of generality it can be  assumed that the state is localized about the origin: an appropriate unitary transformation, generated by  the ${\mathfrak{su}}(2)$, maps a state localized about an arbitrary point to the state, localized about the origin, leaving both the generalized purity and the number of the GCS in the expansion invariant.
 For definiteness let us assume that the $P$-distribution has a finite support area $\cal S$ of radius $\alpha$ about an origin on the phase space. Using the expression for the resolution of identity in terms of the GCS \cite{Gilmore,viera} $\left|\tau\right\rangle$
\begin{eqnarray}
\Op I=\frac{2j+1}{\pi}\int\frac{d^2 \tau}{(1+|\tau|^2)^2}\left|\tau\right\rangle\left\langle \tau \right| \label{iresol}
\end{eqnarray}
the number of the GCS in the expansion of the state can be estimated as follows
\begin{eqnarray}
M[\psi]=\frac{2j+1}{\pi}\int_{\cal S}\frac{d^2 \tau}{(1+|\tau|^2)^2}=(2j+1)\int_0^{|\alpha|^2}\frac{d |\tau|^2}{(1+|\tau|^2)^2}=(2j+1)\frac{|\alpha|^2}{1+|\alpha|^2}. \label{nconf}
\end{eqnarray}
To calculate the generalized purity we must calculate the expectation values of $\Op J_x$, $\Op J_y$ and $\Op J_z$. Given the $P$-representation of the state, the expectation value of an observable $\Op X$ can be calculated using its $Q$-representation:
\begin{eqnarray}
\left\langle \Op X\right\rangle=\frac{2j+1}{\pi}\int \frac{d^2 \tau}{(1+|\tau|^2)^2}P(\tau)Q_{\Op X}(\tau), \label{expectation}
\end{eqnarray}
where $Q_{\Op X}(\tau)=\left\langle \tau\left|\Op X\right|\tau \right\rangle$. We have \cite{Gilmore,viera}
\begin{eqnarray}
Q_{\Op J_x}&=&j\frac{\tau+\tau^*}{1+|\tau|^2}, \nonumber \\
Q_{\Op J_y}&=&j\frac{\tau-\tau^*}{i(1+|\tau|^2)}, \nonumber \\
Q_{\Op J_z}&=&j\frac{|\tau|^2-1}{1+|\tau|^2}. \label{qrep}
\end{eqnarray}
Assuming that $P(\tau)$ is symmetric about the origin ($\tau=0$), we see that the expectation values of $\Op J_x$ and $\Op J_y$ vanish and 
\begin{eqnarray}
\left\langle \Op J_z\right\rangle=\frac{2j+1}{\pi}\int \frac{d^2 \tau}{(1+|\tau|^2)^2}P(\tau)j\frac{|\tau|^2-1}{1+|\tau|^2}. \label{epjz}
\end{eqnarray}
We assume that 
\begin{eqnarray}
P(\tau)=\left\{ \begin{array}{cc} p, \ \ \ \ |\tau|\le |\alpha|  \\ 0,  \ \ \  \ |\tau|>|\alpha|. \end{array} 
 \right. \label{prep}
\end{eqnarray}
The distribution (\ref{prep}) as it stands does not correspond to a pure state. Nonetheless, it can be understood as a coarse grained version of a localized pure state, useful for calculation of the expectations of $\Op J_x$, $\Op J_y$ and $\Op J_z$ and the generalized purity $P_{{\mathfrak{su}}(2)}[\psi]$, Eq.(\ref{gpur}). In fact, Eq.(\ref{qrep}) gives the characteristic scale of unity for the change of the $Q$ representation in the integral (\ref{expectation}). On the other hand, the resolution of identity (\ref{iresol}) implies the characteristic scale of the fine structure of the $P$-distribution (the width of the  overlap of two coherent states)  of the order of $(1+|\tau|^2)/\sqrt{j}$. Therefore, as long as $(1+|\alpha|^2)/\sqrt{j}\ll 1$ in Eq.(\ref{prep}) the coarse grained distribution can be used for calculation of the generalized purity. As can be seen below, Eq.(\ref{gpur}), for $j \gg 1$  the coarse grained description is valid for calculation of the generalized  purity asymptotically as $1/j$.

For a particular form of the distribution  (\ref{prep}),  Eq.(\ref{epjz}) simplifies to
\begin{eqnarray}
\left\langle \Op J_z\right\rangle&=&p \ j (2j+1)\int_0^{|\alpha|^2}\frac{d |\tau|^2}{(1+|\tau|^2)^2}\frac{|\tau|^2-1}{1+|\tau|^2}\nonumber \\
&=&j-p \ j (2j+1)\int_0^{|\alpha|^2}\frac{2 d |\tau|^2}{(1+|\tau|^2)^3}=j-p \ j (2j+1)\left(1-\frac{1}{(1+|\alpha|^2)^2}   \right). \label{epjz1}
\end{eqnarray}
The number $p$ in Eq.(\ref{prep}) can be found from the normalization condition:
\begin{eqnarray}
1&=&\left\langle \Op I\right\rangle=\frac{2j+1}{\pi}\int \frac{d^2 \tau}{(1+|\tau|^2)^2}P(\tau)=p \  (2j+1)\int_0^{|\alpha|^2}\frac{d |\tau|^2}{(1+|\tau|^2)^2}\nonumber \\
&=&p \  (2j+1)\frac{|\alpha|^2}{1+|\alpha|^2} ,
\end{eqnarray}
from which $p=(1+|\alpha|^2)/(|\alpha|^2(2j+1))$. Inserting this expression into Eq.(\ref{epjz1}), we obtain
\begin{eqnarray}
\left\langle \Op J_z\right\rangle&=&j-p \ j (2j+1)\left(1-\frac{1}{(1+|\alpha|^2)^2}   \right)=j-\frac{1+|\alpha|^2}{|\alpha|^2} \ j \left(1-\frac{1}{(1+|\alpha|^2)^2}  \right)\nonumber \\
&=&-\frac{j}{1+|\alpha|^2}. \label{epjz2}
\end{eqnarray}
Therefore,
\begin{eqnarray}
P_{{\mathfrak{su}}(2)}[\psi]=\frac{1}{j^2}\sum_i\left\langle \Op J_i\right\rangle^2=\frac{1}{j^2}\left\langle \Op J_z\right\rangle^2=\left(\frac{1}{1+|\alpha|^2}\right)^2  \label{gpur}
\end{eqnarray}
and
\begin{eqnarray}
|\alpha|^2=\frac{1}{\sqrt{P_{{\mathfrak{su}}(2)}[\psi]}}-1
\end{eqnarray}
Inserting the latter expression into Eq.(\ref{nconf}), we obtain for the number of GCS in the expansion:
\begin{eqnarray}
M[\psi]=(2j+1)\left(1-\sqrt{P_{{\mathfrak{su}}(2)}[\psi]}\right). \label{nconf1}
\end{eqnarray}
As argued after the Eq.(\ref{prep}) expressions (\ref{gpur}) and (\ref{nconf1}) are valid for $P_{{\mathfrak{su}}(2)}[\psi] \gg 1/j$.

\section{Classicality condition: (i) subalgebra ${\mathfrak{su}}(n)$ of single particles observables of the $n$-modes BEC in an optical lattice; (ii) subalgebra of local observables of a system of $n$ $d$-level system. }
\subsection{BEC}

The spectrum-generating algebra of the Bose-Hubbard model for the $n$-modes BEC in optical lattice is ${\mathfrak{su}}(n)$ subulgebra of the single particles observables \cite{tikhonenkov, trimborn}.
It is shown the the classicality condition (\ref{classicality}) is satisfied in this case, provided the number of atoms $N$ in the condensate complies with 
\begin{eqnarray}
N \gg n. \label{becnclass}
\end{eqnarray}
The Hilbert space of the condensate is a totally symmetric irreducible representation of the ${\mathfrak{su}}(n)$  $[N]$ \cite{iachello06} and the value of the Casimir in this representation is \cite{iachello06}
\begin{eqnarray}
 c_{\cal H}=\frac{n-1}{2n}N (N+n).\label{totalcas}
\end{eqnarray}
The total uncertainty in the GCS by \cite{delbourgo, klyachko}
\begin{eqnarray}
\Delta_{min}=c_{\cal H}-\left\langle \Lambda_N| \Lambda_N \right\rangle = c_{\cal H} - \frac{n-1}{2n}N^2= \frac{1}{2}N(n-1) , \label{totalbec}
\end{eqnarray}
where we have used the known expression \cite{iachello06} for the norm of the maximal weight vector \cite{gilmorebook} $\Lambda_N$ in the totally symmetric irreducible representation of the ${\mathfrak{su}}(n)$  $[N]$.
The value of the Casimir in the adjoint representation is \cite{iachello06}
\begin{eqnarray}
 c_{\texttt{adj}}=n . \label{totalcasadj}
\end{eqnarray}

Thus Eq.(\ref{classicality})  holds if and only if Eq.(\ref{becnclass}) holds. Moreover,
\begin{eqnarray}
 \sqrt{\frac{ c_{\cal H}}{c_{\texttt{adj}}}}=\sqrt{\frac{n-1}{2n^2}N (N+n)},
\end{eqnarray}
which  implies Eg.(\ref{interval}), provided Eq.(\ref{becnclass}) holds. 

Therefore, using the stochastic NLS Eq.(\ref{snlse}), propagation can be advantageous for calculation of the single particles observables, provided the on-site interaction preserves the time scale separation in Eg.(\ref{interval}).
\subsection{Local observables}
Let $\mathfrak{g}$ be a subalgebra of local observables on the composite Hilbert space. For simplicity, let us  consider $n$ $d$ level systems in the Hilbert space ${\cal H}=\otimes_{i=1}^n {\cal H}_i$ and  a subalgebra of local observables $\mathfrak{g}=\oplus_{i=1}^n \mathfrak{su}(L)\subseteq \oplus_{i=1}^n \mathfrak{su}(d)\subseteq \mathfrak{su}(d^n)$. Since the minimum of the total uncertainty (\ref{giu}) for a local subalgebra is obtained in a product state $\psi_{prod}=\otimes_{i=1}^n \psi_i$, where each $\psi_i$ is a GCS with respect to the local subalgebra $\mathfrak{su}(L)$ it follows that
\begin{eqnarray}
\Delta_{min}&=&\Delta[\psi_{prod}]=c_{\cal H}-P_{\mathfrak{g}}[\psi_{prod}]=\sum_{i=1}^n\left( c_{{\cal H}_i}-P_{\mathfrak{su}(L)}[\psi_i] \right)=n\left( c_{{\cal H}_d}-P_{\mathfrak{su}(L)}[GCS] \right)\nonumber \\
&=& n\Delta_{d,min}, \label{delmin}
\end{eqnarray}
where ${\cal H}_d$ is the Hilbert space of a $d$-level subsystem and $\Delta_{d,min}$ is the minimal total uncertainty of a state of any subsystem with respect to the subsystem subalgebra $\mathfrak{su}(L)$.
 Therefore, the condition (\ref{classicality}) is equivalent to 
\begin{eqnarray}
\frac{\Delta_{min}}{c_{\cal H}}=\frac{\Delta_{d,min}}{c_{{\cal H}_d}}\ll 1, \label{classicalitybeak}
\end{eqnarray}
i.e., holds if and only if the local subalgebras $\mathfrak{su}(L)$ of the subsystems operators comply with the classicality condition.  For example in the composite system of two-level system the only subalgebra of local observables is the local subalgebra  $\mathfrak{g}=\oplus_{i=1}^n \mathfrak{su}(2)$. The eigenvalue of the local Casimir equals $(1/2)(1/2+1)=3/4$ and the generalized purity with respect to a $\mathfrak{su}(2)$ algebra of each two-level system is $1/4$. Therefore  the minimal total uncertainty with respect to a $\mathfrak{su}(2)$ algebra of each two-level system equals $3/4-1/4=1/2$ and the ratio of the uncertainty to the Casimir equals $(1/2)/(3/4)=2/3$. Therefore, the strong inequality (\ref{classicalitybeak}) is not satisfied. More generally, it can be shown using Eq.(\ref{delmin}) that the local algebra $\mathfrak{g}= \oplus_{i=1}^n \mathfrak{su}(d)\subseteq \mathfrak{su}(d^n)$ gives
\begin{eqnarray}
\frac{\Delta_{min}}{c_{\cal H}}=\frac{d}{d+1}, 
\end{eqnarray}
therefore the classicality condition (\ref{classicality}) does not hold.

\section{Estimation of the number of stochastic realizations, necessary to converge the expectation values of the observables in $\mathfrak g$ to a prescribed absolute accuracy $\epsilon$.}
Given a random variable $\Op X$ with dispersion $D_X\equiv \left\langle \Op X^2\right\rangle-\left\langle \Op X\right\rangle^2$ the number of samplings $n(\epsilon)$, necessary  to estimate the expectation value $\left\langle \Op X\right\rangle$ to the absolute accuracy $\epsilon$ equals 
\begin{eqnarray}
n(\epsilon)_X=\frac{D_X}{\epsilon^2} \label{estimate}.
\end{eqnarray}
Let us assume that each observable $\Op X_i\in \mathfrak g$ is measured in an experiment to a prescribed accuracy $\epsilon$. 
The corresponding number of experimental runs is $n(\epsilon)_{X_i}$. Then
\begin{eqnarray}
 \sum_{i=1}^{\texttt{dim} \{ \mathfrak g\}}n(\epsilon)_{X_i}&=&\frac{\sum_{i=1}^{\texttt{dim} \{ \mathfrak g\}} D_{X_i}}{\epsilon^2}=\frac{\sum_{i=1}^{\texttt{dim} \{ \mathfrak g\}} \left(\left\langle \Op X_i^2\right\rangle-\left\langle \Op X_i\right\rangle^2\right)}{\epsilon^2}\nonumber \\
&=& \frac{ C_{\cal H}-\sum_{i=1}^{\texttt{dim} \{ \mathfrak g\}}\left\langle \Op X_i\right\rangle^2}{\epsilon^2}\label{estimate1}.
\end{eqnarray}

Now consider the computation of expectation values of observables $\Op X_i\in \mathfrak g$ in a state $\Op \rho(t)$, evolving according to the  Eq.  (\ref{liouville}), by averaging over stochastic unravelings (\ref{snlse}). By Eq.(\ref{estimate}) the number of unravelings necessary to compute the expectation value of $\Op X_i$  to the accuracy $\epsilon$ is $n(\epsilon)_{X_i}'=D_{X_i}'/\epsilon^2$, where $D_{X_i}'$ is the dispersion of the observable in the state $\Op \rho(t)$. Then
\begin{eqnarray}
 \sum_{i=1}^{\texttt{dim} \{ \mathfrak g\}}n(\epsilon)_{X_i}'&=&\frac{\sum_{i=1}^{\texttt{dim} \{ \mathfrak g\}} D_{X_i}'}{\epsilon^2}=\frac{\sum_{i=1}^{\texttt{dim} \{ \mathfrak g\}} \left(\left\langle \Op X_i^2\right\rangle'-\left\langle \Op X_i\right\rangle'^2\right)}{\epsilon^2}\nonumber \\
&=& \frac{ C_{\cal H}-\sum_{i=1}^{\texttt{dim} \{ \mathfrak g\}}\left\langle \Op X_i\right\rangle'^2}{\epsilon^2}\label{estimate2},
\end{eqnarray}
where $<\Op X>'$ means statistical average over the unravelings of the quantum expectation values obtained in each unraveling (which is the random variable for the purpose of Eq.(\ref{estimate})). But on the  time interval of the simulation (Sec.III)
\begin{eqnarray}
\left\langle \Op X_i\right\rangle'=\left\langle \Op X_i\right\rangle, \label{equalexp}
\end{eqnarray}
therefore Eqs.(\ref{estimate1}),(\ref{estimate2}) and (\ref{equalexp}) imply
\begin{eqnarray}
\sum_{i=1}^{\texttt{dim} \{ \mathfrak g\}}n(\epsilon)_{X_i}'=\sum_{i=1}^{\texttt{dim} \{ \mathfrak g\}}n(\epsilon)_{X_i}.
\end{eqnarray}
It follows that 
\begin{eqnarray}
 n(\epsilon)_{st} &\equiv & \texttt{max}_i \{ n(\epsilon)_{X_i}'\} \le \sum_{i=1}^{\texttt{dim} \{ \mathfrak g\}}n(\epsilon)_{X_i}'=\sum_{i=1}^{\texttt{dim} \{ \mathfrak g\}}n(\epsilon)_{X_i}\le \texttt{dim} \{ \mathfrak g\} \texttt{max}_i \{ n(\epsilon)_{X_i}\}\nonumber \\
 &\equiv& \texttt{dim} \{ \mathfrak g\} n(\epsilon)_{ex},
\end{eqnarray}
where $n_{ st}(\epsilon)$ is the number of stochastic realizations, necessary to obtain the expectation value of \textit{each} observable $\Op X_i \in \mathfrak g$ to an absolute accuracy $\epsilon$, $n_{ ex}(\epsilon)$ is the number of experimental runs, necessary to obtain the expectation value of \textit{each} $\Op X_i$ to the absolute accuracy $\epsilon$.

 \end{document}